\documentclass[twocolumn,showpacs,preprintnumbers,amsmath,amssymb]{revtex4}

\usepackage{graphicx}
\usepackage{dcolumn}
\usepackage{bm}

\newcommand{\Vtip}{V_{\textrm{tip}}}
\newcommand{\Vbias}{V_{\textrm{bias}}}
\newcommand{\Vcpd}{V_{\textrm{cpd}}}
\newcommand{\Gdot}{G_{\textrm{dot}}}
\newcommand{\IQPC}{I_{\textrm{QPC}}}
\newcommand{\Phiin}{\Phi^{\textrm{in}}}
\newcommand{\trans}{\textrm{d}I/\textrm{d}V_{\textrm{dg}}}
\newcommand{\const}{\textrm{const.}}
\newcommand{\eq}{\begin{equation}}
\newcommand{\qe}{\end{equation}}

\begin{document}

\title{Imaging a Coupled Quantum Dot - Quantum Point Contact System}

\author{A. E. Gildemeister}
\email{gildemeister@phys.ethz.ch}
\author{T. Ihn}
\author{R. Schleser}
\author{K. Ensslin}

\affiliation{Laboratory of Solid State Physics, ETH Z{\"u}rich,
CH-8093 Z{\"u}rich, Switzerland}

\author{D. C. Driscoll}
\author{A. C. Gossard}

\affiliation{Materials Department, University of California,
Santa Barbara, CA-93106, USA}

\date{13 February 2007}

\begin{abstract}
We performed measurements on a quantum dot and a capacitively coupled quantum point contact by using the sharp metallic tip of a low-temperature scanning force microscope as a scanned gate. The quantum point contact served as a detector for charges on the dot or nearby. It allowed us to distinguish single electron charging events in several charge traps from charging events on the dot. We analyzed the tip-induced potential quantitatively and found its shape to be independent of the voltage applied to the tip within a certain range of parameters. We estimate that the trap density is below 0.1\% of the doping density and that the interaction energy between the quantum dot and a trap is a significant portion of the dot's charging energy. Possibly, such charge traps are the reason for frequently observed parametric charge rearrangements.
\end{abstract}


\maketitle


\section{Introduction}

Quantum point contacts (QPCs) and quantum dots are two basic building blocks of semiconductor nanostructures. Their characteristic features are the conductance quantization of QPCs \cite{Wees:1988} and the charge quantization in quantum dots \cite{Kouwenhoven:1997} observed in clean samples at low temperatures. A QPC can be tuned to a regime where it is sensitive to single charges in its vicinity. In this way electrons entering or leaving a nearby quantum dot can be detected \cite{Field:1993}. Scanning gate measurements, where the conductive tip of a low-temperature scanning force microscope (SFM) is scanned over the sample surface at constant height, have been reported for both QPCs and quantum dots. Single electron charging of quantum dots \cite{Woodside:2002,Pioda:2004,Fallahi:2005} and charge detection with QPCs \cite{Crook:2002,Pioda:2007} have been addressed individually.

In Ref.~\cite{Woodside:2002} quantum dots forming in carbon nanotubes could be located and single electron charging was investigated. Similar results were obtained for quantum dots prepared in a GaAs/AlGaAs heterostructure \cite{Pioda:2004}, in which the single-electron regime could be realized \cite{Fallahi:2005}. Some scanning gate measurements of QPCs have been interpreted in terms of charge detection~\cite{Crook:2002,Pioda:2007} while others focus on quantum interference effects \cite{Cunha:2006} as they have previously been reported to be detectable outside of QPCs \cite{Topinka:2001}.

We have used scanning gate microscopy to study the combined system of a quantum dot and a QPC similar to the one used in Ref.~\cite{Field:1993}. We employed the QPC as a charge-readout for the quantum dot and as a sensor for other charges in its proximity. The quantum dot allowed us to gauge the tip potential and to make a quantitative analysis.


\section{Experimental Setup}

\begin{figure}[htb]
\includegraphics[width=8.2cm]{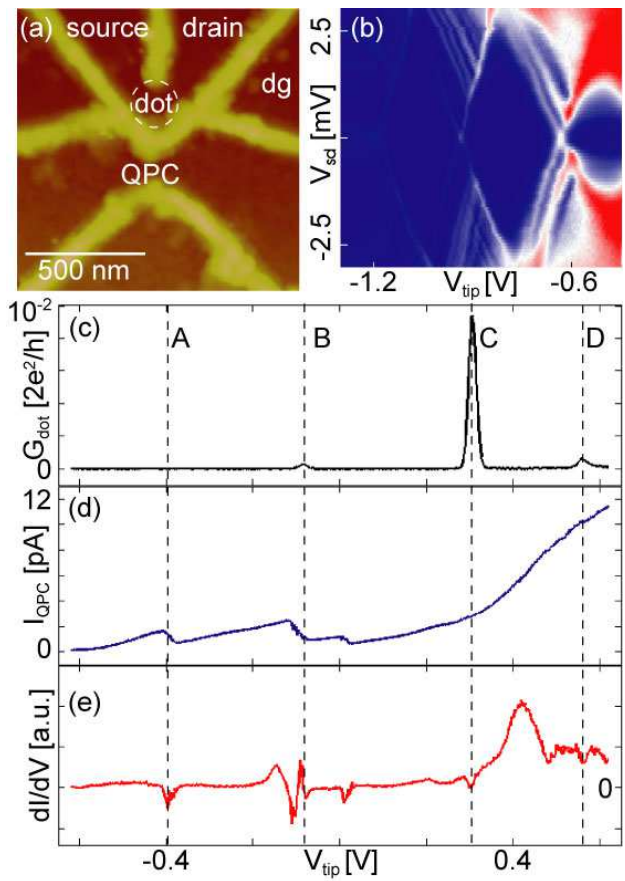}
\caption{\label{fig:traces}  (Color online) (a) Room temperature scanning force microscope image of the quantum dot and the QPC formed by the bright oxide lines. The drain gate is labeled dg. (b) Charge stability diagram where the dot conductance is shown as a function of source-drain bias and tip voltage. Here the tip was at a constant position over the center of the dot. (c) The quantum dot conductance $\Gdot$, (d) QPC current $\IQPC$, and (e) QPC transconductance $\trans$ as functions of the tip voltage $\Vtip$ for the tip positioned over the center of the dot.}
\end{figure}

For the experiments presented here we used a GaAs/AlGaAs heterostructure  with a two-dimensional electron gas (2DEG) residing 34~nm below the surface to prepare our sample. The mobility was about 450'000~cm$^2$/Vs and the electron density was 4$\times 10^{11}$~cm$^2$ at 4.2~K. Local anodic oxidation with a room-temperature SFM \cite{Held:1997} was used to define a quantum dot with a geometrical diameter of about 150~nm and an adjacent QPC. The structure is depicted in Fig.~\ref{fig:traces}(a).

To record a scanning gate image, the PtIr tip of a low-temperature SFM was scanned over the sample surface at a constant height of about 200~nm \cite{tipheight}  and  the conductances of the quantum dot and the QPC were spatially mapped. We applied a voltage $\Vtip$ of a few hundred millivolts between the tip and the 2DEG. The SFM was operated in a dilution refrigerator \cite{Gildemeister1:2007} and the electronic temperature in the experiment presented here was about 500~mK. We used standard lock-in techniques to measure conductances.


Using lateral gates, we adjusted the quantum dot to the Coulomb blockade regime, as it can be seen in the charge stability diagram in Fig.~\ref{fig:traces}(b) where the tip was positioned over the center of the dot and used as a plunger gate. While the quantum dot was in Coulomb blockade, we tuned the QPC to a conductance below its first quantized plateau. Here it is very sensitive to small changes of the surrounding electrostatic potential. Removing one electron from the dot would, for example, increase the QPC conductance. We also measured the QPC transconductance by applying a small AC voltage of 0.5 mV to the  drain gate on the opposite side of the quantum dot and detecting the QPC conductance at the same frequency. This measures the transconductance, i.e., the derivative of the QPC conductance with respect to the drain gate voltage, and is a sensitive technique to detect charging of a quantum dot. However, the QPC is of course sensitive to all nearby charging events.

With the tip positioned over the center of the dot we measured, as a function of the voltage $\Vtip$ applied to the tip, the conductance of the dot $\Gdot$, the QPC current $\IQPC$, and the QPC transconductance $\trans$, as shown in Figs.~\ref{fig:traces}(c-e). We can discern three peaks labeled B,C,D in $\Gdot$ [Fig.~\ref{fig:traces}(c)] as single electrons are loaded onto the dot. At the same positions in $\Vtip$  we find, as expected, three dips of similar magnitude in $\trans$ [Fig.~\ref{fig:traces}(c)] whereas the change in slope of $\IQPC$ [Fig.~\ref{fig:traces}(b)] at these tip voltages is more difficult to see. The dip in $\trans$ labeled A could be caused by another resonance in the dot that does not show up in the dot current because the measurement is not sensitive enough. The remaining structure in $\IQPC$ and $\trans$ is not due to the dot and will be discussed below. We purposely present typical data where charging events from sources other than the dot can be distinguished. While for a single tip position one can find a set of parameters where no charging events outside the dot influence the measurement, it is unavoidable that such events influence measurements in which the tip is scanned.


\section{Scanning Gate Measurements}

\begin{figure*}[tb]
\includegraphics[width=17cm]{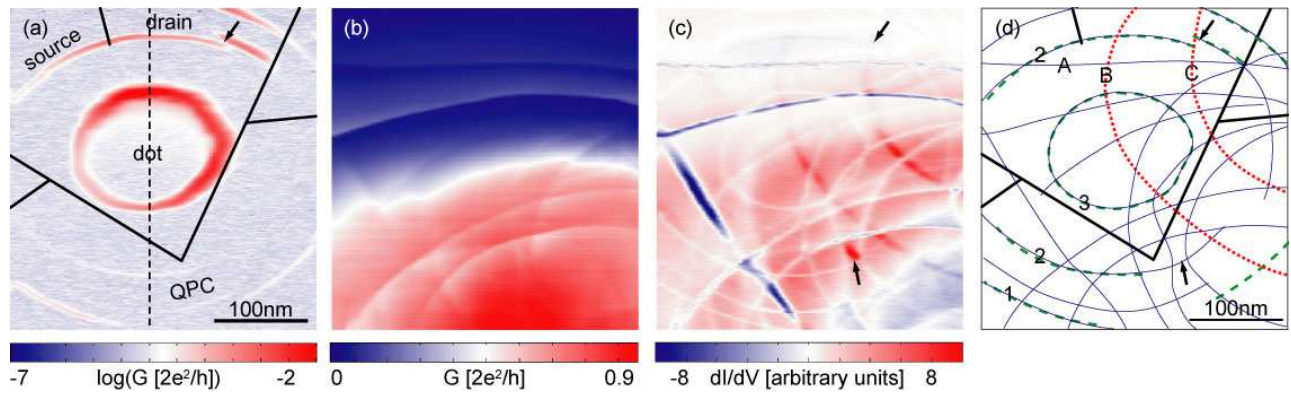}
\caption{\label{fig:gdotgqpcdidu} (Color online) (a-c) Scanning gate measurements of the coupled quantum dot and QPC system that were recorded simultaneously. (a) Conductance of the quantum dot. The black lines indicate the approximate position of the oxide lines that define the structure. The dashed line shows where the measurements of Figs. \ref{fig:diamondsandpotential}, \ref{fig:vdependent}, and \ref{fig:height} were recorded. (b) Conductance of the QPC. (c) Transconductance of the QPC. (d) Overlay of rings from (a) (dashed green lines) and  arcs from (c) (solid blue for dips and dotted red for peaks). The labels are explained in the main text. Similar data has been presented in Ref. \cite{Gildemeister5:2007}.}
\end{figure*}

In Fig.~\ref{fig:gdotgqpcdidu} we show three scanning gate images  that were simultaneously measured with a constant tip voltage of $\Vtip = 425$~mV in a 15 hour long scan. In Fig.~\ref{fig:gdotgqpcdidu}(a) we see the dot conductance, in Fig.~\ref{fig:gdotgqpcdidu}(b) the QPC conductance, and in Fig.~\ref{fig:gdotgqpcdidu}(c) the QPC transconductance. These measurements were remarkably reproducible and for a given set of parameters we observed no time-dependence of the result.

For most tip positions we see no conductance of the dot because it is in Coulomb blockade. The near-circular ring of high conductance in the center occurs when the tip-induced potential brings a quantized state of the dot in resonance with the electrochemical potential of source and drain. Two more concentric circles of high conductance can be seen partly. From identical measurements at different tip voltages $\Vtip$ we know that the tip-induced potential is attractive, so that single electrons are added to the dot as the tip moves closer to its center \cite{Gildemeister1:2007}. The rings are similar in diameter as those shown in Refs. \cite{Fallahi:2005,Kicin:2005,Gildemeister1:2007} and the smallest rings in Ref. \cite{Woodside:2002}, whereas in Ref. \cite{Pioda:2004} only much larger rings are shown. The rings are much narrower than in either of the previous reports except those in Ref. \cite{Gildemeister1:2007} which were recorded during the same cooldown.

The conductance of the QPC [Fig.~\ref{fig:gdotgqpcdidu}(b)] increases as the attractive tip comes closer. Additionally, we see ring-shaped kinks. These are more pronounced in the QPC transconductance [Fig.~\ref{fig:gdotgqpcdidu}(c)] where we can distinguish about 15 rings or arcs. In Fig.~\ref{fig:gdotgqpcdidu} (d) we manually traced the rings on the dot conductance (dashed lines, green online) and the transconductance [solid (blue online) for dips and dotted (red online) for peaks] and combined them in a single graph.

For every ring of high dot conductance we find a corresponding ring in the QPC transconductance: These are charging events on the dot. Of the remaining arcs some are centered roughly at the position of the QPC: These could be transmission resonances that occur at given potentials in the QPC. We are left with several arcs that are centered neither in the dot nor the QPC: We interpret these as the signatures of charge traps in the vicinity of the QPC \cite{Pioda:2007}.


\section{Tip-Induced Potential}

\begin{figure}[bt]
\includegraphics[width=8cm]{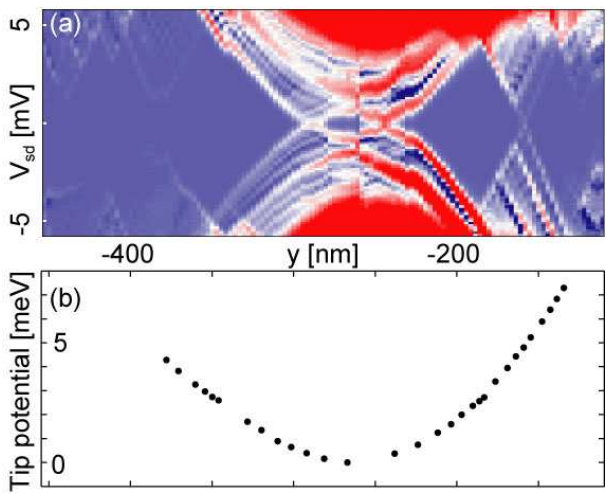}
\caption{\label{fig:diamondsandpotential} (Color online) (a) The dot conductance as a function of source-drain bias and the $y$-position of the tip measured along the dashed line in Fig.~\ref{fig:gdotgqpcdidu}(a). By mapping the edges of the Coulomb diamonds it is possible to deduce (b) the tip-induced potential energy $-e\Phi^0(y)$.}
\end{figure}

A quantitative analysis of these arcs requires a quantitative understanding of the tip potential, i.e., the potential that the tip induces in the sample. In two separate measurements we investigated the shape and magnitude of the potential as well as its origin. In order to analyze this potential we have used the quantum dot as a very sensitive detector for the electric potential. 

We determined the tip potential in $y$-direction along the dashed line through the dot center that is shown in Fig.~\ref{fig:gdotgqpcdidu}(a). We moved the tip stepwise along this line and measured the dot conductance as a function of source-drain bias  for every point. The result is the charge stability diagram shown in Fig.~\ref{fig:diamondsandpotential}(a). This plot differs from standard Coulomb diamond measurements in that here the horizontal axis represents the tip position rather than a gate voltage. Nonetheless this plot can be used to determine fundamental properties such as the dot's charging energy. In particular, the tip potential can be read off this charge stability diagram. The tip works like a plunger gate and its potential shifts the energy of charge states in the dot. For a given tip position the shift in energy can be determined from the bias voltage that is necessary so that the state comes within the bias window and can contribute to the conductance.  This happens at the onset of conductance at the edge of the Coulomb diamonds. By following the edge of a Coulomb diamond we can determine the energy of a charge state in the dot, and thereby the tip potential, as a function of the tip position. The method can be used for a sequence of diamonds by appropriately changing the sign and adding offsets to the bias voltage values read. The resulting tip potential is shown in Fig.~\ref{fig:diamondsandpotential}(b). Note that the bias voltage values read from the charge stability diagram are a factor of two higher than the tip potential because we applied $+\Vbias/2$ to the source and $-\Vbias/2$ to the drain. We have multiplied the electrostatic tip potential with the electron charge, i.e. $-e$, to plot the more intuitive potential energy.

We see only the central part of the tip potential which has an approximately parabolic shape. We expect the potential to be bell-shaped and to become flat when the tip is moved far away from the dot as it can be inferred from Ref. \cite{Pioda:2004}. The charge stability diagram was measured for a voltage $\Vtip^0=200$~mV and hence the tip potential $\Phi^0(y)$ was determined for this particular tip voltage.

\begin{figure}[tb]
\includegraphics[width=8.2cm]{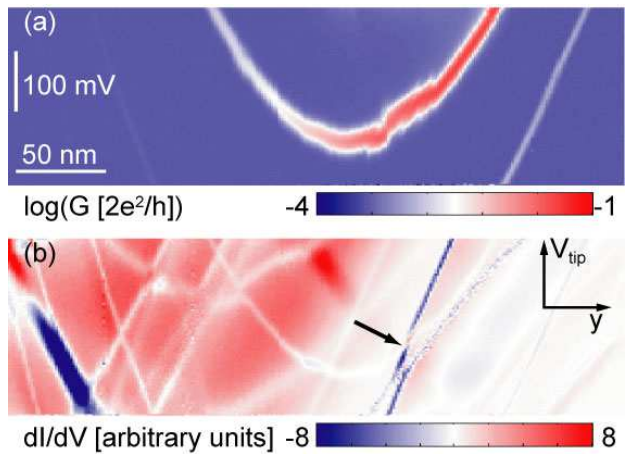}
\caption{\label{fig:vdependent} (Color online) (a) Dot conductance as a function of the voltage $\Vtip$ applied to the tip and the $y$-position of the tip, measured along the dashed line in Fig.~\ref{fig:gdotgqpcdidu}(a). (b) The transconductance of the QPC measured simultaneously with the dot conductance. The arrow marks an anticrossing due to capacitive coupling between the dot and a charge trap. }
\end{figure}

In order to better understand the properties of the tip potential we have again moved the tip stepwise along the same line. Now the source-drain bias was around zero and we swept the voltage $\Vtip$ applied to the tip from 125~mV to 425~mV. In Fig.~\ref{fig:vdependent}(a) we show the resulting zero-bias dot conductance as a function of $\Vtip$ and the $y$-position. We see two resonances and can read off $\Vtip(y)$, i.e., the voltage we need to apply to the tip so that a particular charge state of the dot remains in resonance with the Fermi levels of the leads.

We invoke a general electrostatic model \cite{Kicin:2005, Hallam:1996} for the dot to understand the connection between $\Phi^0(y)$ and $\Vtip(y)$. We write $\alpha(y)$ for the $y$-dependent lever-arm of the tip, $\Vtip$ for the voltage applied to the tip, $\Vcpd$ for the contact potential difference between tip and 2DEG and $\Phiin (y)$ for a portion of the tip potential that is independent of $\Vtip$. A tip-voltage independent potential has been observed before in Refs. \cite{Woodside:2002} and \cite{Pioda:2004}. For the electrochemical potential $\mu_N$ of the quantum dot with $N$ electrons we write
$$
\mu_N = \const - e \alpha(y) \Vtip + \underbrace{e \alpha(y) \Vcpd
- e \tilde{\Phi}^{\textrm{in}}(y)}_{-e\Phiin(y)},
$$
assuming that the lever-arms of the in-plane gates are independent of tip position and tip voltage. The complex physics of the quantum dot and the effect of the in-plane gates are subsumed in the constant term. 

With the Coulomb diamonds we have measured
\eq
\mu_N(y) = \Phi^0(y) =\const -e \alpha(y)\Vtip^0 - e \Phiin(y),
\label{eq:diamonds}
\qe
where we also assume that source and drain have very similar lever arms. This is fulfilled, as can be seen from the symmetry of the Coulomb diamonds with respect to zero bias in Fig.~\ref{fig:traces}(b) and Fig.~\ref{fig:diamondsandpotential}(a). 

For $\Vtip(y)$ in Fig.~\ref{fig:vdependent}(a) we see that
\eq
\mu_{N^{\prime}} = \const -e \alpha(y)\Vtip(y) - e \Phiin(y).
\label{eq:Vofy}
\qe
We choose a point $y_0$ as a reference so that, if we consider only differences, we can use Eqs.~(\ref{eq:diamonds}) and (\ref{eq:Vofy}) to calculate
\eq
\alpha(y)=\frac{\Phi^0(y) - \Phi^0(y_0)}{\Vtip(y) - \Vtip(y_0)}
+\alpha(y_0)\frac{\Vtip(y_0) - \Vtip^0}{\Vtip(y) - \Vtip^0}
\label{eq:alpha}
\qe
For $y_0$ at the center of the dot we have measured the additional charge stability diagram shown in Fig.~\ref{fig:traces}(b) in which we swept $\Vtip$ for a fixed tip position. 
From this we find $\alpha(y_0) \approx 1.1\%$.
Since $\Vtip(y_0) \approx \Vtip^0$, the second term in Eq. (\ref{eq:alpha}) vanishes and since both $\Phi^0(y)$ and $\Vtip(y)$ can be well approximated by a parabola, the first term is constant in $y$. 

Therefore we find that in the range of about $\pm 75$~nm around the potential minimum where we can measure it, $\alpha(y)$ is constant within about 10\%.  The absolute value is very close to that found in Ref. \cite{Kicin:2005} where $\alpha$ was also found to vary little over comparable length scales. Calculating the first term of Eq.~(\ref{eq:alpha}) also leads to $\alpha \approx 1.1 \%$, corroborating the value from the independent measurement of $\alpha(y_0)$. If $\alpha$ is constant then the spatial variation of the tip potential can be regarded as independent of the voltage applied to the tip, i.e. $\Phi^0(y) = \alpha(y_0) \Vtip^0 + \Phiin (y)$. Clearly, for larger distances between tip and dot the tip's lever arm $\alpha$ should approach zero. This behavior can be qualitatively seen in Ref. \cite{Pioda:2004}. We have measured the tip potential only along one line through the dot center but since in Fig.~\ref{fig:gdotgqpcdidu}(a) we see near-circular lines we assume that the tip potential has circular in-plane symmetry.


\section{Analysis of Scanning Gate Measurements}

With this quantitative understanding of the tip potential at hand we can further analyze the scanning gate images of Fig.~\ref{fig:gdotgqpcdidu}. For instance, the innermost ring of high dot conductance is wider than the rings further away from the center and the outer rings are more closely spaced. Both effects are due to the increasing steepness of the tip potential with increasing tip-dot distance.

We interpret the arcs seen in the QPC transconductance in Fig.~\ref{fig:gdotgqpcdidu}(c) that are not centered around the dot or the QPC as an effect of charge traps. Most arcs in the QPC transconductance are of similar strength and width as those associated with single electron charging events on the dot and their centers are located at a similar distance from the QPC as the dot. Therefore we attribute them to single charges added or removed from charge traps.  Similar images can be seen in Refs. \cite{Pioda:2007} and \cite{Aoki:2006} and possibly in Ref. \cite{Cunha:2006}. One may wonder what the influence of charge traps on scanning gate images of other nanostructures, such as rings \cite{Hackens:2006}, could be. The positions of the traps are at the center of the arcs. Most of the traps are close to the dot with the exception, for example, of the trap corresponding to the ring labeled ``A" in Fig.~\ref{fig:gdotgqpcdidu} (d), which has a small curvature, suggesting that the trap could be several microns away.

From the visible charge traps we crudely estimate the trap density to be of the order of $n_t \approx 30$~$\mu$m$^{-2}$, that is, below $0.1\%$ of the doping density of $n=120'000$~$\mu$m$^{-2}$ which is about ten times lower than the value estimated in Ref. \cite{Pioda:2007}. Obviously, this value will depend strongly on the quality of the sample. We can also estimate the average distance between two traps to be $d_t \approx 2\sqrt{\pi/n_t} \approx 600$~nm. This distance is of the same order of magnitude as the quantum scattering length which is typically a factor of 10 shorter than the mean free path \cite{Coleridge:1989} which was $\ell = 4.7\ \mu$m in this sample. We see that even for a sample of good quality with relatively high mobility there is a high chance of having charge traps in close proximity to any given structure.


Anticrossings of different rings show the capacitive interaction between traps and the dot [upper arrows in Figs.~\ref{fig:gdotgqpcdidu} (a), (c), and (d) at the intersection of the rings labeled 2 and C] or in between traps [lower arrows in Figs.~\ref{fig:gdotgqpcdidu} (c) and (d)]. Other examples of anticrossings can be seen in Fig.~\ref{fig:vdependent}(b).
We focus on the first example because the energy scales of the dot are known for this configuration. We will first discuss how the charge state of the trap changes and then estimate the interaction energy.

To the right of C the addition of an electron to the dot occurs 
for a larger distance between tip and dot than to the left of C. As we have found the tip potential with respect to the dot to be positive, this implies that to the right of C there is one positive charge more on the trap than to the left of C. While the dot becomes more negatively charged as the tip approaches, the trap becomes more positively charged when the tip comes closer. Presumably, this is also the reason why in the transconductance rings 1, 2, and 3 are dips and C is a peak. 

We can determine the interaction energy $\Delta E$ between dot and trap from the width of the gap in ring 2 together with the known tip potential and find $\Delta E \approx 1$~meV. This is a substantial portion of the dot's charging energy of about 5~meV [cf. Fig. \ref{fig:diamondsandpotential}(a)]. In a simple Coulomb interaction model we can estimate the distance $d$ between dot and trap from $\Delta E$. We take into account that the dot is very close to the surface of the Ga[Al]As sample with its high dielectric constant of $\epsilon=12.8$, a configuration that creates an image charge of equal sign \cite{Jackson:1974}. We do not take into account the screening effect of the 2DEG because the 2DEG is mostly depleted by the oxide lines between the dot and this particular trap. We estimate 
$$
d \approx \frac{e^2}{4 \pi \epsilon_0 }\frac{2}{1+\epsilon}\frac{1}{\Delta E} \approx 200\ \textrm{nm,}
$$
which is only a little bit less than the distance between the dot and the center of ring C. Presumably, it is this capacitive coupling between a system under study and charge traps around it which causes the well-known parametric charge rearrangements that often impair data quality. An example of such impairment is the extra structure in $\IQPC$ and $\trans$ in Fig. \ref{fig:traces}(d-e) that is not due to the dot.


\section{Effect of the tip height}

\begin{figure}[t]
\includegraphics[width=8.2cm]{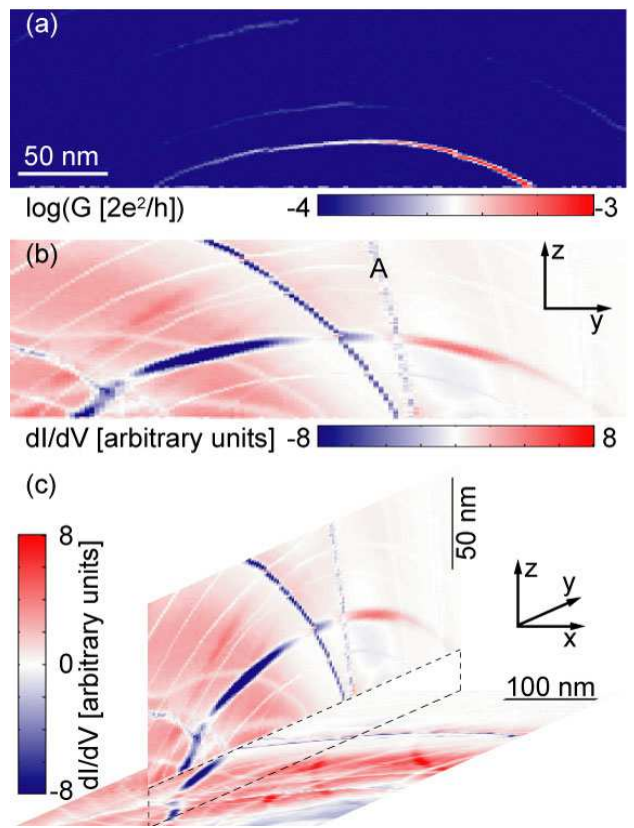}
\caption{\label{fig:height} (Color online) (a) The dot conductance and (b) the QPC transconductance as a function of the vertical distance between tip and sample and the $y$-position of the tip, measured along the dashed line in Fig.~\ref{fig:gdotgqpcdidu}(a). (c) Three-dimensional representation of the QPC transconductance.}
\end{figure}

We have also measured the behavior of the dot and the QPC as a function of the vertical distance between tip and sample. This method could potentially help to determine the depth of a charge trap or the tip potential in $z$-direction. The dot conductance and the QPC transconductance are shown in  Fig.~\ref{fig:height} (a) and (b) while Fig.~\ref{fig:height} (c) shows a three-dimensional representation of the QPC transconductance. It visualizes equipotential surfaces of the tip with respect to the dot and the traps. This measurement does help to localize traps in the $x-y$ plane and, for example, confirms that the trap corresponding to the arc labeled A is particularly far away from the dot. However, the data were insufficient for a more quantitative analysis.


\section{Conclusion}

In conclusion we have presented scanning gate measurements of a coupled quantum dot - QPC system. By measuring the dot and QPC conductance and the QPC transconductance we could identify and locate several charge traps. By moving the tip, these traps could be charged with single electrons. This led to reproducible arcs in the transconductance measurements which were constant in time. Using an unconventional Coulomb diamond measurement we characterized and gauged the tip potential and found that, within the investigated range of parameters, its shape did not depend on the voltage applied to the tip. We found the trap concentration to be very low compared to the doping density while the interaction energy between a trap and the dot could be a significant portion of the dot's charging energy. We suggest that the charge traps could be the reason for the well-known parametric charge rearrangements.

For future scanning gate experiments the exact shape of the tip potential needs to be investigated further. Metallic gates on top of the sample would help to avoid the effects of the charge traps by screening them from the tip potential.

We acknowledge financial support from ETH Z{\"u}rich.



\begin{thebibliography}{20}
\bibitem[{tip({The tip height was constant throughout the measurements. Due to
  technical limitations it could not be determined exactly.})}]{tipheight}
 \bibinfo{year}{{The tip height was constant throughout the measurements. Due
  to technical limitations it could not be determined exactly.}}
\end{thebibliography}

\end{document}